\documentclass[twocolumn,showpacs,prl,amsmath,amssymb]{revtex4}
\usepackage{graphicx}     
\begin{document}

\title{Theory of Tunneling in the Exciton Condensate 
of Bilayer Quantum Hall Systems}
\author{K. Park}
\affiliation{Condensed Matter Theory Center, 
Department of Physics, University of Maryland,
College Park, MD 20742-4111}
\date{\today}       

\begin{abstract}
We develop a theory of interlayer tunneling in the exciton condensate
of bilayer quantum Hall systems, which predicts
strongly enhanced, but finite, 
tunneling conductance peaks near zero bias 
even at zero temperature.
It is emphasized that, though this strongly enhanced tunneling 
originates from spontaneous interlayer phase coherence, 
it is fundamentally not the Josephson effect.
Because of strong interlayer correlation, the bilayer system behaves
as a single system so that conventional tunneling theories
treating two layers as independent systems are not applicable.
Based on our theory, we compute the height of conductance peak
as a function of interlayer distance, 
which is in good agreement with experiment.
\end{abstract}

\maketitle

When Spielman {\it et al.} \cite{Spielman} observed 
strongly enhanced interlayer conductance peaks near zero bias
in bilayer quantum Hall systems at 
total filling factor $\nu_T=1$,
they not only renewed our interest 
in the bilayer quantum Hall effect \cite{Murphy},
but also attracted intense interest 
from the general perspective of strongly correlated physics.
It was because,
in addition to its many-body origin,
the bilayer quantum Hall effect bears a rather precise analogy
to superconductivity;
the ground state of bilayer quantum Hall effect at 
interlayer distance $d/l_B \ll 1$ ($l_B=\sqrt{\hbar c/eB}$)
maps onto the BCS wavefunction of an exciton condensate
of particle-hole pairs formed across the interlayer barrier.
In fact, Bose-Einstein condensation of excitons
in semiconductors has been sought after for decades. 
In particular, there have been fascinating
recent experiments on the possible condensation of 
optically generated indirect excitons \cite{Indirect}, 
for which, however, there is not yet conclusive evidence.
On the other hand, it is generally accepted that
the strongly enhanced conductance peak in the quantum Hall regime
is a direct indication of macroscopic phase coherence.

To be concrete regarding the mapping between 
the superconductivity and bilayer quantum Hall effect, 
let us write the exact ground state wavefunction at $d/l_B=0$,
{\it i.e.} the Halperin's (1,1,1) state \cite{Halperin111}
(which is adiabatically connected to the ground states 
at sufficiently small, but finite $d/l_B$):
\begin{equation}
|\psi_{111}\rangle = \prod_{m} 
(c^{\dagger}_{m\uparrow} + c^{\dagger}_{m\downarrow}) |0\rangle ,
\label{111}
\end{equation}
where $m$ is a momentum index in the lowest Landau level and 
the pseudospin representation is used:
$\uparrow$ ($\downarrow$) indicates the top (bottom) layer.
Note that Eq.(\ref{111}) describes the full wavefunction including
both orbital and layer degree of freedom \cite{Yang}.
Since Eq.(\ref{111}) has a structure isomorphic to the BCS wavefunction, 
it is clear that the bilayer quantum Hall state should have
a phase coherence between states with different interlayer number
difference in analogy with phase coherence between different
number eigenstates in superconductivity, 
which is the origin of the Josephson effect. 
Naturally,
this similarity led previous authors \cite{WenZee,Ezawa}
to predict the Josephson effect
in bilayer quantum Hall systems.
The strongly enhanced conductance observed by Spielman {\it et al.},
therefore, seemed to be exactly the experimental verification needed.
However, there are key properties of the conductance peak indicating that
this phenomenon is not the conventional Josephson effect:
most notably, saturation of height as well as width  
to finite values in the limit of zero temperature \cite{Spielman2}.

This apparent discrepancy gave rise to two groups of thought. 
In one group, the enhanced conductance is still regarded as 
DC Josephson effect, but its height is reduced 
by complicated disorder-induced 
fluctuations \cite{Stern1,Stern2,Balents,Fogler}.
On the other hand, others \cite{Joglekar} argued that
there is no exact analog of Josephson effect
in interlayer tunneling experiments 
because the bilayer system as a whole is a single superfluid,
not a set of two superfliud systems.
While we agree with the latter viewpoint 
that the enhanced interlayer tunneling conductance
is not the analog of Josephson effect, 
we show below that strong interlayer correlation requires
a fundamentally new starting point different from all of above theories
in order to construct a self-consistent theory of interlayer tunneling
in quantum Hall regime.

As mentioned previously, the bilayer quantum Hall system
is a single superfluid system. So, it is impossible to
induce a chemical potential gradient between the two layers
without destroying interlayer phase coherence, 
in which case the interlayer current becomes a normal current,
not supercurrent.
It is important to distinguish between 
the chemical potential gradient and applied interlayer bias voltage 
because, even when the bias voltage is applied, 
bilayer systems will immediately reach an equilibrium
by creating charge imbalance in order to compensate 
the relative voltage difference and therefore there is no
chemical potential gradient.
Though this point seems straightforward,
it has been completely overlooked
by all previous theories which, 
regardless of their viewpoint 
regarding the analogy with Josephson effect,
began by implicitly making a self-contradictory assumption
that there is strong interlayer correlation due to 
the Coulomb interaction
but two layers can be treated independently by
having a finite chemical potential gradient. 
In fact, if one can induce a finite chemical potential gradient 
while maintaining interlayer phase coherence, there would be a very
interesting experimental consequence: oscillating tunneling
current whose frequency is proportional to the applied bias voltage.
However, no oscillating current has been observed in experiments.

\begin{figure}
\includegraphics[width=3.5in]{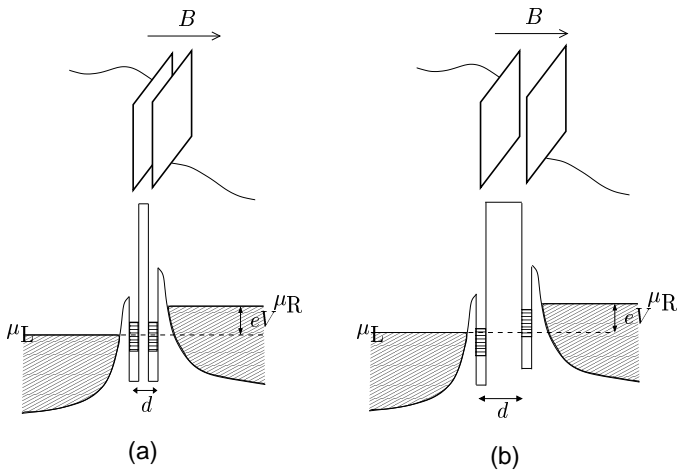}
\caption{Schematic diagram of tunneling measurement in bilayer quantum 
Hall systems.
Note that there is no chemical potential gradient between layers when
the ground state of bilayer system becomes a single exciton condensate at 
small interlayer distance $d$, as depicted in (a).
A consistent theory of interlayer tunneling, therefore,
should inevitably take
external leads into consideration.
On the other hand, when $d$ is sufficiently large as shown in (b),
two layers behave as independent systems, 
and interlayer coherence is lost. 
\label{fig1}}
\end{figure}

Now, if there is no interlayer chemical potential gradient, 
there is no electromotive force within bilayer system and 
any current should be induced from outside.
It is, therefore, necessary to take into account external leads,
as schematically shown in Fig.\ref{fig1}.
This, of course, makes any quantitative prediction
dependent on the way in which bilayer systems are connected to
external leads.
However, it is still possible to make a quantitative prediction on
essential aspects of coherent interlayer tunneling. 
In particular,
we will compute the dependence of tunneling conductance peak height
on interlayer distance $d/l_B$. 
Also, we will show that the width
is finite even at zero temperature, and it is controlled
ultimately by extremely small, 
but finite single-particle interlayer tunneling gap 
$\Delta_{\textrm{SAS}}$.

Let us begin our quantitative analysis
by writing the total Hamiltonian including
the Hamiltonian for Coulomb interaction 
between electrons in bilayer system $H_0$, 
the Hamiltonian describing the left and right lead, 
$H_L$ and $H_R$ respectively, and
tunneling between leads and the bilayer system $H'$:
\begin{eqnarray}
H &=& H_0 + H' +H_R +H_L,
\end{eqnarray}
\begin{eqnarray}
\frac{H_0}{e^2/\epsilon l_B} &=&
{\cal P}_{LLL} \Big(
\sum_{i,j\in\uparrow} \frac{1}{r_{ij}}
+\sum_{k,l\in\downarrow} \frac{1}{r_{kl}}
\nonumber \\
&+&\sum_{i\in\uparrow,k\in\downarrow}
\frac{1}{\sqrt{r^2_{ik}+(d/l_B)^2}}
\Big) {\cal P}_{LLL},
\\
H' &=& \sum_{k,m} T_{R\uparrow}(k,m)
[c^{\dagger}_R(k)c_{m\uparrow} + \textrm{h.c.}]
\nonumber \\
&+& \sum_{p,m'} T_{L\downarrow}(p,m')
[c^{\dagger}_L(p)c_{m'\downarrow} + \textrm{h.c.}],
\end{eqnarray}
where, as before, the pseudospin representation is used, and
${\cal P}_{LLL}$ is the lowest Landau level projection operator.
$T_{R\uparrow}(k,m)$ is the tunneling amplitude 
between the state with momentum $k$ in the right lead, and 
the state with $m$ in the top layer of bilayer system.
$T_{L\downarrow}(p,m')$ is similarly defined.
$H_R$ and $H_L$ describe
electrons in external leads as normal Fermi liquids.
It is now very important to note that
$H$ does not have any interlayer tunneling term within the bilayer system. 
It is because we are interested in 
the spontaneous interlayer coherence 
which occurs in the limit of zero interlayer tunneling gap: 
$\Delta_{\textrm{SAS}}/(e^2/\epsilon l_B) \rightarrow 0$.
As will be shown later, this spontaneous interlayer coherence 
is due to the many-body effect of Coulomb interaction in $H_0$, and
it creates a non-zero current from one layer to the other 
even in the limit of zero interlayer tunneling gap
(of course, in unbiased equilibrium, the net current is zero 
since two opposite currents cancel each other).

Since there is no direct process of transporting
electrons from one lead through the bilayer system to the other lead,
one has to consider second order tunneling processes:
\begin{equation}
H'_T = H' \frac{1}{E_g-H_0-H_R-H_L} H',
\end{equation}
where $E_g$ is the ground state energy of
$H_0+H_R+H_L$.
By adding an electron to the top layer and removing another
from the bottom layer, $H'_T$ describes
tunneling processes through the bilayer system.
Now, because the bilayer quantum Hall state is incompressible at
sufficiently small $d/l_B$, 
adding or removing electrons
costs a finite energy which is equal to
the Coulomb self-energy of quasi-particles, 
$\Delta_C$ \cite{comment_Delta_C}. 
We will compute $\Delta_C$ as a function of $d/l_B$
later by using exact diagonalization.
It is, however, sufficient at this stage to know that
$\Delta_C$ is independent of momentum $m$ in the lowest Landau level.
So one can just replace $H_0+H_R+H_L-E_g$ by $\Delta_C$.
Remember that there is
no energy cost in taking electrons from external leads 
because normal Fermi liquids are compressible.

Now, we assume that the tunneling amplitudes 
$T_{R\uparrow}(k,m)$ and $T_{L\downarrow}(p,m')$
are more or less independent of momenta $k$ and $p$,
which is a common practice in tunneling theories 
when studying tunneling processes only within a narrow region
of energy near Fermi surface.
Keeping only terms of $H'_T$ relevant for transporting electrons
from one lead to the other, we arrive at the following
tunneling Hamiltonian:
\begin{equation}
H_T = \sum_{k,p} \left[
c^{\dagger}_{R}(k)c_L(p) \hat{T}^{\dagger}
+c^{\dagger}_{L}(p)c_R(k) \hat{T}
\right],
\label{H_T}
\end{equation}
where 
\begin{equation}
\hat{T}=\frac{1}{\Delta_C}\sum_{m} T_{RL}(m) 
c^{\dagger}_{m\uparrow} c_{m\downarrow}
\end{equation}
and $T_{RL}(m)=T_{R\uparrow}(k_F,m)T_{L\downarrow}(k_F,m)$ \cite{comment_m}.
Based on $H_T$, the tunneling current operator $\hat{J}$
is given as follows:
\begin{equation}
\hat{J} = e i \sum_{k,p} \left[
c^{\dagger}_{R}(k)c_L(p) \hat{T}^{\dagger}
-c^{\dagger}_{L}(p)c_R(k) \hat{T}
\right].
\label{J}
\end{equation}

\begin{figure}
\includegraphics[width=1.7in]{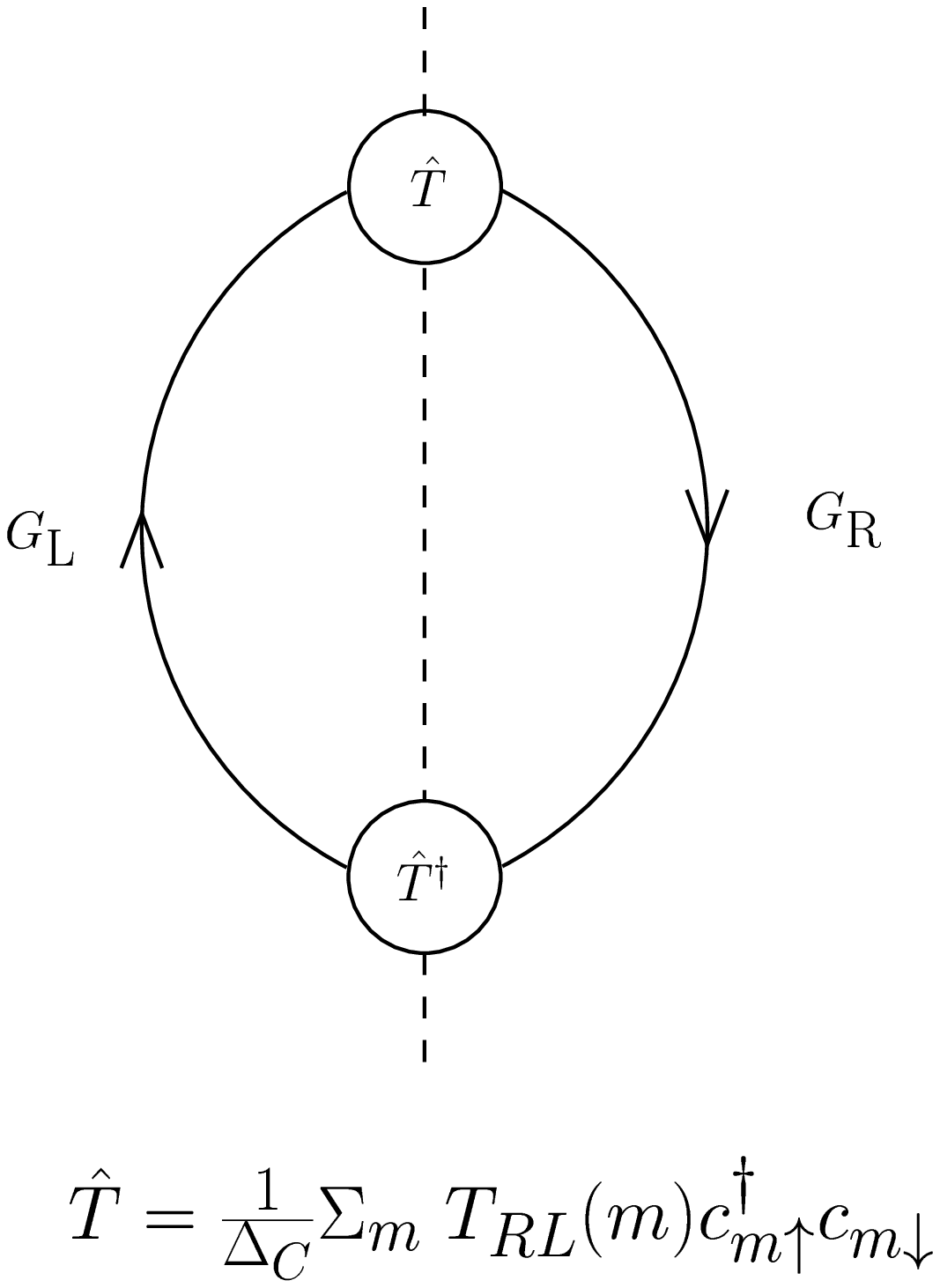}
\caption{Feynman diagram of interlayer tunneling 
in bilayer quantum Hall systems. 
The vertex operator $\hat{T}$
contains all of many-body effects of an exciton condensate.
$T_{RL}$ is the tunneling amplitude 
and $\Delta_C$ is the Coulomb self-energy of quasiparticle.
\label{fig2}}
\end{figure}

We now compute the expectation value of current operator
via a conventional first-order S-matrix expansion:
\begin{equation}
I(t)= -i \int^{t}_{-\infty} d t' 
\langle [\hat{J}(t), H_T(t')] \rangle .
\label{current}
\end{equation}
The new aspect of our tunneling theory is 
the vertex operator $\hat{T}$
which contains all of many-body effects of the exciton condensate.
Eq.(\ref{current}) can be evaluated further
using the Feynman diagram depicted in Fig.\ref{fig2}:
\begin{eqnarray}
I &=& 2e |\langle\hat{T}\rangle|^2 \sum_{k,p}
\int^{\infty}_{-\infty} \frac{d \varepsilon}{2\pi}
A_R(k,\varepsilon)A_L(p,\varepsilon+eV)
\nonumber \\
&\times&[f(\varepsilon)-f(\varepsilon+eV)]
\nonumber \\
&=& 4\pi e^2 D_R D_L |\langle\hat{T}\rangle|^2 V
\label{Ohm}
\end{eqnarray}
where $A_R$ ($A_L$) is the spectral function of the right (left) lead,
$f(\varepsilon)$ is  the usual Fermi-Dirac distribution function, and
$D_R$ ($D_L$) is the density of states at the Fermi surface of
right (left) lead.
It is clear from Eq.(\ref{Ohm}) that there is no DC Josephson effect
because the conductance 
$G$ ($\equiv dI/dV \propto |\langle\hat{T}\rangle|^2$) 
is finite.
However, the interlayer tunneling current is
zero unless there is a phase coherence: $\langle\hat{T}\rangle \neq 0$.
Remember that $\langle\hat{T}\rangle$ measures  
a phase coherence
between states with different values of interlayer number difference, 
$N_{\textrm{rel}}$,
because $\hat{T} \propto c^{\dagger}_{m\uparrow} c_{m\downarrow}$
and therefore changes $N_{\textrm{rel}}$ by two.
So, unless the ground state is a coherent
linear combination of states with various $N_{\textrm{rel}}$,
$\langle\hat{T}\rangle$ is zero, and so is the tunneling current.
As mentioned before,
this is similar to the phase coherence  
between different number eigenstates in superconductivity, 
which is responsible for the Josephson effect.
In this sense, interlayer tunneling conductance 
is related to the Josephson effect. However,
we emphasize that the conductance  
should be finite
even at zero temperature and there is no direct analogy with
the Josephson effect.
We now compute the interlayer tunneling conductance
as a function of $d/l_B$. 
In particular, we will be interested in normalized conductance
since the absolute scale of conductance
is sensitive to sample-specific details such as $D_R$, $D_L$ and
$T_{\textrm{RL}}$. 

\begin{figure}
\includegraphics[width=2.0in,angle=-90]{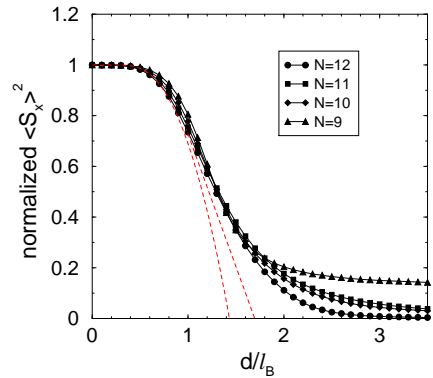}
\caption{Normalized expectation value of condensate order parameter 
${\langle S_x \rangle}^2$. 
Dashed lines indicate the upper and lower bound
for an estimate of the thermodynamic limit of ${\langle S_x \rangle}^2$
as a function of $d/l_B$.
$N$ is the total number of electrons.
\label{fig3}}
\end{figure}

In essence, we compute $|\langle\hat{T}\rangle|^2$ 
which can be further reduced as follows:
\begin{equation}
|\langle\hat{T}\rangle|^2 =
\frac{\langle S_x \rangle^2}{\Delta^2_C} 
\left| \frac{1}{N} \sum_m T_{\textrm{RL}}(m) \right|^2 ,
\label{T2}
\end{equation}
where $N$ is the total number of electrons, and
we have used the fact that 
$\langle c^{\dagger}_{m\uparrow} c_{m\downarrow} \rangle$
is independent of $m$ and is equal to $\langle S_x \rangle /N$.
$S_x$ 
[$= \sum_m 
(c^{\dagger}_{m\uparrow} c_{m\downarrow} 
+c^{\dagger}_{m\downarrow} c_{m\uparrow})/2$]
is the order parameter of exciton condensation,
and it can also be interpreted as the pseudospin magnetization 
in the $x$ direction.
Since $\sum_m T_{\textrm{RL}}(m)/N$ does not depend on $d/l_B$,
the interlayer distance dependence of conductance is solely 
determined by $\langle S_x \rangle^2/\Delta^2_C$.

In Fig.\ref{fig3} we plot $\langle S_x \rangle^2$
as a function of $d/l_B$
which is computed via exact diagonalization of 
finite systems with various particle numbers in torus geometry.
When computing $\langle S_x \rangle$ in finite systems,
it is very important 
to take into account fundamental fluctuations in $N_{\textrm{rel}}$; 
the true ground state is a coherent, linear combination of
states with various $N_{\textrm{rel}}$ \cite{Park,Park2}.
Though estimating the accurate thermodynamic limit 
of $\langle S_x \rangle^2$ is difficult,
it is reasonable to argue that
the true thermodynamic limit lies between
two dashed lines in Fig.\ref{fig3}.

\begin{figure}
\includegraphics[width=2.0in,angle=-90]{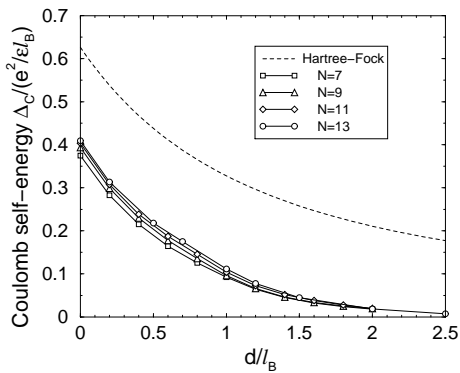}
\caption{Coulomb self-energy of a quasiparticle.
\label{fig4}}
\end{figure}

Fig.\ref{fig4} plots the Coulomb self-energy of a quasiparticle,
$\Delta_C/(e^2/\epsilon l_B)$, as a function of $d/l_B$ which is determined
in exact diagonalization studies
by computing the energy gap of particle-hole-pair 
excitation with the largest momentum and taking half of its value.
For comparison, we also plot
the self-energy in the Hartree-Fock approximation \cite{Fertig}
which tends to overestimate $\Delta_C$.

\begin{figure}
\includegraphics[width=2.0in,angle=-90]{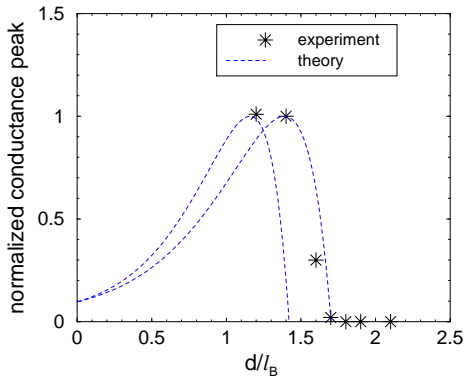}
\caption{Normalized interlayer tunneling conductance peak 
as a function of interlayer distance 
in comparison with experimental data from Ref.\cite{Spielman}.
We define the normalized conductance 
as conductance divided by its maximum value 
as a function of $d/l_B$. 
Two theoretical curves are obtained from the upper and lower 
bound of thermodynamic estimate in Fig.\ref{fig3}.
\label{fig5}}
\end{figure}

Finally, in Fig.\ref{fig5} we compare our estimate of
normalized interlayer tunneling conductance near zero bias,
{\it i.e.} $\langle S_x \rangle^2/\Delta^2_C$,
with experimental data of Spielman {\it et al.} \cite{Spielman}.
We define the normalized conductance 
as conductance divided by its maximum value 
as a function of $d/l_B$. 
Two dashed lines in Fig.\ref{fig5}
correspond to the upper and lower bound of estimated thermodynamic limits
of $\langle S_x \rangle^2$ in Fig.\ref{fig3}. 
Considering simplifications used in our theory
such as omission of finite thickness effect,
we find our theory to be in good agreement with experiments.
In addition to further comparison with experiments
in the regime $d/l_B \gtrsim 1.2$,
it will be very interesting to see whether
our prediction of decrease of conductance peak 
for $d/l_B \lesssim 1.2$ is consistent with 
future experiments.
Remember that decrease in conductance peak at small $d/l_B$
is due to increase in energy gap to put electrons
into bilayer systems while the pseudospin magnetization
is saturated.
We would like to emphasize that, once normalized,  
our theoretical estimate of conductance peak
does not have any fitting parameter.

We have shown by means of Eq.(\ref{Ohm}) and (\ref{T2}) 
that in exciton condensate
the interlayer tunneling conductance at small bias is 
finite, but strongly enhanced.
However, we did not show why the conductance should be sharply
peaked near zero bias, which we will explain now.
Once the interlayer current is driven by an external electromotive force, 
it should physically flow through
the bilayer system since otherwise there is no steady state.
Exciton condensates accomplish this by adjusting 
their interlayer phase difference $\phi$ to sustain 
the externally driven current,
which is again easy to understand in terms of 
the ground state wavefunction at $d/l_B \rightarrow 0$:
\begin{equation}
|\psi_{111}(\phi)\rangle = \prod_{m} 
(c^{\dagger}_{m\uparrow} + e^{i\phi} c^{\dagger}_{m\downarrow})
|0\rangle ,
\label{111phi}
\end{equation}
which carries a net internal current within bilayer system
equal to
$e\Delta_{\textrm{SAS}}\frac{N}{2}\sin{\phi}$ \cite{comment_current}.
Then, there should be a critical current at $\phi=\pi/2$ 
which is the maximum current allowed without breaking phase coherence.
Therefore, for sufficiently large voltage bias,
coherent interlayer currents should be cut off 
and become constant as a function of bias voltage,
once they reach the critical value controlled by 
single-particle interlayer tunneling gap $\Delta_{\textrm{SAS}}$.
The conductance associated with coherent tunneling, therefore, should
be zero after the critical voltage and  
is strongly enhanced only near zero bias.
Consequently,
the width of conductance peak is proportional to
very small, but finite $\Delta_{\textrm{SAS}}$,
while the proportionality constant
strongly depends sample-specific details 
such as the density of states of leads.
It is, however, encouraging to find that typical width of conductance peak
($\sim 10-100 \mu eV$)
is roughly in the same order as $\Delta_{\textrm{SAS}}$ 
\cite{Spielman,Spielman2}.
The above argument is valid for general $d/l_B$
when there is phase coherence.

Until now, we have studied the interlayer tunneling conductance
in a single bilayer system, which, we showed, is not 
the exact analog of Josephson effect. 
We now conclude by proposing a much more direct analog
with the Josephson effect.
Consider a pair of bilayer systems, say A and B (four layers altogether),
separated by a lateral tunneling barrier.
Then, put an interlayer current through the top and bottom layer
of, say, bilayer system A, in which way a non-zero interlayer phase
difference is induced in bilayer system A while the system B has none.
We predict then that there will be two counterflowing currents:
one between two, top layers of system A and B, and 
the other between bottom layers.
The net current will be zero, but it may be possible to measure these
two currents individually.

This work was supported by ARDA.



\begin{thebibliography}{}
\bibitem{comment_Delta_C} Bilayer quantum Hall systems have
an edge excitation which requires an infinitesimally small energy
to be created.
However, the edge excitation does not have any contribution
to coherent interlayer tunneling since 
it is not a part of coherent exciton condensate
and its contribution to interlayer tunneling
is only through single-particle tunneling
which is of the order of extremely small single-particle tunneling gap
$\Delta_{\textrm{SAS}}$.
\bibitem{comment_m} In general, 
the vertex operator $\hat{T}$ can have contributions
from terms $c^{\dagger}_{m\uparrow} c_{m'\downarrow}$
with $m \neq m'$. However, it has been shown \cite{Park}
that the ground state at sufficiently small $d/l_B$
has very little contribution due to 
$\langle c^{\dagger}_{m\uparrow} c_{m'\downarrow} \rangle$
when $m \neq m'$.
\bibitem{comment_current} This can be shown 
by computing the expectation value of internal current operator
within bilayer systems:
$\hat{J}_{\textrm{inter}}=ie\Delta_{\textrm{SAS}} \sum_m 
(c^{\dagger}_{m\uparrow} c_{m\downarrow} 
-c^{\dagger}_{m\downarrow} c_{m\uparrow})/2$
which is basically the pseudospin component in the $y$ direction
similar to $S_x$.
\end{thebibliography}
\end{document}